\documentclass[letterpaper, 10 pt, conference]{ieeeconf}  

\IEEEoverridecommandlockouts                              

\usepackage[pdftex]{graphicx}
\usepackage{amsmath} 

\usepackage{amsfonts}
\usepackage{algorithm}
\usepackage{algorithmic}
\usepackage{lipsum}
\usepackage{subcaption}

\usepackage{amssymb}
\usepackage{xcolor}
\usepackage[colorlinks=true, urlcolor=blue, linkcolor=red]{hyperref}
\usepackage{ulem}
\usepackage[most]{tcolorbox}
\usepackage{multirow}
\usepackage{gensymb}
\usepackage{array}
\usepackage{ragged2e}
\usepackage{caption}
\usepackage{adjustbox}
\usepackage[T1]{fontenc}

\title{\LARGE \bf
Anti-Sensing: Defense against Unauthorized Radar-based Human Vital Sign Sensing with Physically Realizable Wearable Oscillators}

\author{Anonymous Authors}
\author{Md Farhan Tasnim Oshim$^{1}$, Nigel Doering$^{2}$, Bashima Islam$^{3}$,  Tsui-Wei Weng$^{2}$, and Tauhidur Rahman$^{2}$
\thanks{*This work is in part supported by the National Science Foundation under grant 2320678 (PI Rahman) and start up grant support from Halıcıoğlu Data Science Institute – UC San Diego, and Manning College of Information \& Computer Sciences – UMass Amherst.}
\thanks{$^{1}$Md Farhan Tasnim Oshim is with Manning College of Information and Computer Sciences, University of Massachusetts Amherst, MA, USA
        {\tt\small farhanoshim@cs.umass.edu}}%
\thanks{$^{2}$Nigel Doering, Lily Weng, and Tauhidur Rahman are with Halıcıoğlu Data Science Institute,
University of California San Diego, CA, USA
        {\tt\small nfdoerin@ucsd.edu, lweng@ucsd.edu, trahman@ucsd.edu}}%
\thanks{$^{3}$Bashima Islam is with the Department of Electrical and Computer Engineering,
Worcester Polytechnic Institute, MA, USA
        {\tt\small bislam@wpi.edu}}%
}

\begin{document}

\maketitle
\thispagestyle{empty}
\pagestyle{empty}

\begin{abstract}

Recent advancements in Ultra-Wideband (UWB) radar technology have enabled contactless, non-line-of-sight vital sign monitoring, making it a valuable tool for healthcare. However, UWB radar's ability to capture sensitive physiological data, even through walls, raises significant privacy concerns, particularly in human-robot interactions and autonomous systems that rely on radar for sensing human presence and physiological functions. In this paper, we present Anti-Sensing, a novel defense mechanism designed to prevent unauthorized radar-based sensing. Our approach introduces physically realizable perturbations, such as oscillatory motion from wearable devices, to disrupt radar sensing by mimicking natural cardiac motion, thereby misleading heart rate (HR) estimations. We develop a gradient-based algorithm to optimize the frequency and spatial amplitude of these oscillations for maximal disruption while ensuring physiological plausibility. Through both simulations and real-world experiments with radar data and neural network-based HR sensing models, we demonstrate the effectiveness of Anti-Sensing in significantly degrading model accuracy, offering a practical solution for privacy preservation.

\end{abstract}

\begin{figure*}
\centering
  \includegraphics[width=0.95\textwidth]{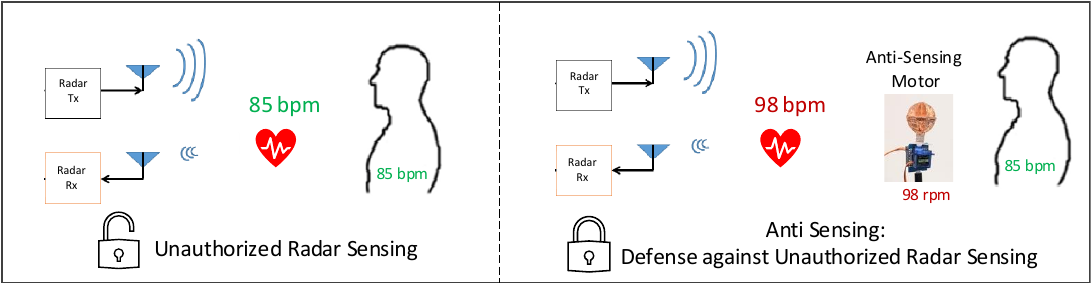}
  \caption{Anti-Sensing Scenario: The left panel shows a radar system detecting an individual's heart rate (85 bpm) without permission. In contrast, the right panel illustrates our anti-sensing solution, where a motor generates a false heart rate signal (98 bpm), effectively blocking the radar from sensing the person's true heart rate (85 bpm) and ensuring privacy.}
  \label{fig:teaser}
\end{figure*}



\section{Introduction}

Recent advancements in contactless sensing technologies, particularly using Ultra-Wideband (UWB) radar, have enabled various applications such as vital signs monitoring \cite{lazaro2010analysis, goldfine2020respiratory, oshim2022eulerian, xie2022deepvs, zhao2020heart, wu2022adaptable} and gesture recognition \cite{ahmed2021uwb, 9253617, 7828517} without physical contact. These technologies offer convenience and efficiency but also raise significant privacy concerns due to their ability to capture sensitive personal information in public spaces without consent. Radar's high penetration capability allows it to sense through walls, posing a potential threat to privacy. In various settings, from homes to public spaces like transportation stops and waiting rooms, it can expose individuals' vital signs, gestures, and behavioral data, leading to the possible misuse of personal information. Unauthorized estimation of vital signs, particularly heart rate, poses significant privacy concerns, as it can disclose sensitive health information, such as stress levels, without consent, leading to discrimination, health profiling, and surveillance, ultimately infringing on personal autonomy and security.

While the benefits of contactless radar sensing are evident, the security and privacy implications remain under-explored. Adversarial attacks, a growing concern in machine learning and computer vision, pose a novel threat to radar-based sensing. Adversarial attacks typically employ white-box methods such as FGSM \cite{goodfellow2014explaining}, IFGSM \cite{alexey2016adversarial}, PGD \cite{mkadry2017towards}, JSMA \cite{papernot2016limitations}, DeepFool \cite{moosavi2016deepfool} , and C\&W \cite{carlini2017towards}, or black-box techniques like ZOO \cite{chen2017zoo}, GenAttack \cite{alzantot2019genattack}, and Boundary Attacks \cite{brendel2017decision, andriushchenko2020square, guo2019simple}. Although most attacks have focused on the digital domain, the exploration of perturbations in the physical domain to deceive such systems remains limited. While some research exists on physical-world attacks in the context of computer vision \cite{eykholt2018robust, duan2021adversarial, li2023physical, huang2023t}, the radar system domain has not been similarly investigated. In particular, the susceptibility of radar data to physically realizable perturbations, such as deliberate, imperceptible modifications to sensor inputs, has not been adequately addressed. These physical perturbations can serve as a defense mechanism for our setting, where they could be used to mislead radar-based sensing systems, resulting in inaccurate vital sign estimations, thereby protecting an individual's privacy.

This paper proposes a novel physically realizable perturbation technique, Anti-Sensing, a pipeline designed to effectively disrupt and deceive unauthorized radar sensing models. Our approach utilizes physical perturbations in the form of oscillating devices, for example, motors as a defense against unauthorized vital sign monitoring, heart rate (HR) in particular. The oscillating frequencies and span of these devices are optimized using a gradient-based defense algorithm that we designed to ensure maximum loss between model predictions and the ground truths. These perturbations mimic legitimate cardiac motion and sufficient noise components, thereby misleading radar-based recognition models into producing inaccurate heart rate predictions. By strategically introducing oscillatory signals that simulate natural human heart rate, our method aims to protect individuals' privacy by thwarting unauthorized radar sensing attempts. We validate our system's success through simulated perturbations on a real dataset and neural network architectures, and subsequently validate it using real collected data with physical devices. Through experiments and analysis, we demonstrate that our proposed perturbation method can effectively deceive radar sensors, making it a practical solution for safeguarding personal privacy in public and private settings. The key contributions of our work can be summarized as follows:
\begin{itemize}
    \item We present Anti-Sensing, a novel perturbation technique that uses oscillatory motions optimized by an appropriate gradient-based attack objective to deceive radar sensing models for regression tasks. This approach introduces a new paradigm in adversarial defense by creating physical perturbations that mimic legitimate human heart rates.
    
    \item We demonstrate that our attacks are physically realizable using oscillating devices, which can be easily worn on the wrist. We use a programmable servo motor to generate variable frequency motion for defense against unauthorized heart rate estimations. These devices allow for precise, reproducible physical perturbations that can effectively counteract radar-based sensing mechanisms.
    
    \item We validate our anti-sensing approach through simulated perturbations applied to a real dataset collected at a sleep clinic and to state-of-the-art neural network-based radar sensing models, followed by testing with data collected using the anti-sensing oscillating device. Our results show that the proposed perturbations are effective in disrupting unauthorized radar sensing models, offering a practical solution for safeguarding individuals' heart rate information.
    
\end{itemize}



\section{Related Works}
Most adversarial perturbations against contactless sensing have primarily focused on the digital domain, involving white-box or black-box attacks on the collected data. For example, Xie et al. \cite{xie2024universal} demonstrated universal targeted adversarial attacks in the digital domain against mmWave-based Human Activity Recognition (HAR), effectively deceiving different models, including voxel-based and heatmap-based, while remaining entirely in the digital domain. Similarly, Ozbulak et al. \cite{ozbulak2021investigating} showed the vulnerability of radar-based CNNs for human activity recognition to both white-box and black-box adversarial attacks, revealing that even minimal perturbations, such as those applied only to input padding, can significantly alter model predictions.

Staat et al. \cite{IRShield} proposed IR-Shield, a countermeasure using intelligent reflecting surfaces (IRSs) to obfuscate wireless channels, achieving detection rates of $5\%$ or less in advanced Wi-Fi-based human motion attacks.
RF-Protect \cite{rfprotect} presents a hardware reflector coupled with a generative mechanism to produce realistic human trajectories aimed at enhancing privacy by introducing artificial human reflections into FMCW radar data to guard against unauthorized through-wall monitoring. However, these existing approaches are ineffective against attacking UWB radar-based sensing, particularly for vital signs, as UWB relies on precise time-of-flight measurements, and unlike FMCW or mmWave systems, delay and frequency in UWB systems are uncorrelated.

\section{Background of Radar-based Sensing}

Ultra-wideband (UWB) radar is a non-invasive sensing technology that emits nanosecond-duration electromagnetic pulses across a broad frequency spectrum (3.1–10.6 GHz), enabling precise detection of both macro-scale movements, such as gestures, and micro-scale physiological activities, including breathing and heartbeats. Its wide bandwidth provides high spatial resolution, making it well-suited for capturing detailed motion data. Additionally, UWB radar’s broad frequency range and pulse characteristics enable non-line-of-sight sensing, allowing detection through walls and other obstacles due to its superior penetration properties. By analyzing the time delay of reflected signals, it tracks a target's movement, while advanced signal processing removes clutter from static objects, ensuring accurate monitoring in dynamic environments - particularly in healthcare, human-object interaction, and human-robot interaction applications.

A received signal at the UWB receiver can be written as the following equation,

\begin{equation}
r(t)=\sum_{j=1}^{L}a_{j}s(t-\tau_{j})+w(t)
\end{equation}

Here, \( s(t) \) is the transmitted pulse, \( a_j \) and \( \tau_j \) represent the amplitude and the propagation delay of the \( j^{\text{th}} \) multipath component, \( w(t) \) is the additive noise from the channel, and \( L \) is the total number of reflected paths. The time delay \( \tau_j \), also known as the time of flight (ToF), can be used to calculate the target's distance \( d_j \) using the relation \( d_j = \frac{c \cdot \tau_j}{2} \), where \( c \) is the speed of light. For simplicity, we will use \( d_j \) in the equation directly to refer to the target distance range bin.

\subsection*{Synthetic Vital Sign Motion Generation}
This subsection demonstrates the process of synthesizing vital sign motion based on ultra-wideband (UWB) radar principles. UWB pulses can be modeled as Gaussian-modulated sinusoidal signals, and a single radar scan of a point target at a fixed range bin \( d_j \) can be expressed as:

\begin{equation}
\small
s(t_i, d_j) = \exp\left(-\left(\frac{t_i - d_j}{\omega_0 / T_s}\right)^2\right) \cdot \exp\left(i \cdot 2\pi \cdot f_0 \cdot T_s (t_i - d_j)\right)
\end{equation}
where, 
\begin{equation}
d_j = \text{A} \cdot \sin\left(2\pi \frac{f_{\text{osc}}}{60 \cdot F_s} \cdot x_i\right) + \text{offsets}[k]
\label{eqn: offset_equation}
\end{equation}

and, 
\begin{itemize}
    \item \( T_s \): Fast-time sampling period
    \item \( \omega_0 \): Width of the Gaussian pulse
    \item \( f_0 \): Frequency of the radar pulse in Fast Time
    \item \( f_\text{osc} \): Frequency of the sinusoidal motion a.k.a. frequency of the Vital Sign (HR) in rpm
    \item \( F_s \): Sampling frequency of the radar in Slow Time.
    \item \(M\): Total number of radar scans
    \item \(N\): Total number of range bins
    \item \(\text{A}\): Spatial amplitude
    \item \(\text{offsets}\): Target locations
    \item \( x_i \): Scan indices from 1 to \(M\)
    \item \( d_j \): Range bin positions modulated by the synthetic vital sign motion from 1 to \(N\).
\end{itemize}

\begin{figure}[htb!]
    \centering
    \includegraphics[width=0.45\textwidth]{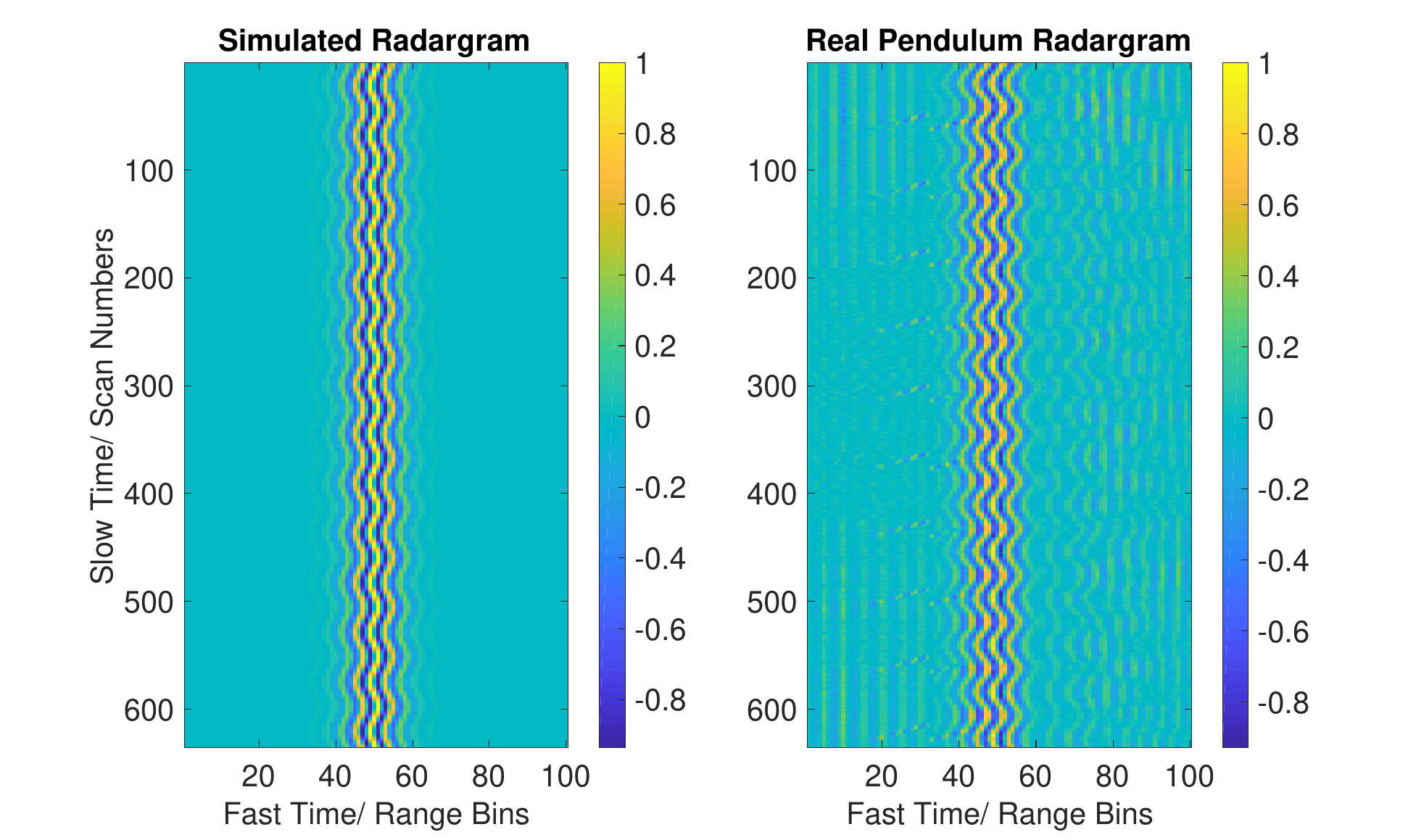} 
    \caption{Comparison of synthetic radargram (left) and real radargram (right) of a point target (a pendulum with a metal bob) oscillating at a frequency of $90$ rpm ($1.5$ Hz).}
    \label{fig: synthetic_radargram}
\end{figure}

Offsets determine the initial target location. If there are multiple targets, synthetic sinusoids are generated for each of them and then summed up together to form a single radar scan. By stacking the Gaussian-modulated pulses over all scans and offsets, we get a 2D radargram for a specific observation window. The $x$-axis of the radargram, also known as the fast-time axis, denotes distance or range, while the $y$-axis, also known as the slow-time axis, indicates time. To demonstrate how closely our simulation matches reality, we present Figure \ref{fig: synthetic_radargram}, a side-by-side comparison of a pendulum with a metal bob oscillating at $90$ rpm ($1.5$ Hz) positioned at one meter from the radar.


\section{Proposed Anti-sensing Method:}

Our goal is to develop a defense mechanism that deceives radar sensing models by introducing deliberate, physically realizable perturbations to radar inputs. Specifically, we aim to mislead radar systems, such as those used for vital sign estimation (e.g., heart rate), into generating inaccurate predictions, thereby protecting individual privacy. This problem is framed as an attack on radar sensing models to induce erroneous estimations. In our proposed algorithm, we optimize oscillatory motions through a gradient-based attack to deceive the vital sign estimation model. Since the perturbations act as a defense mechanism to safeguard privacy, we refer to our algorithm as Sinusoidal Defense Algorithm, a white-box targeted attack requiring knowledge of the vital sign model, as illustrated in Algorithm \ref{alg:sinusoidal_attack}.


A typical white-box classification adversary with a targeted attack optimizes the following objective, where $h_\theta(\cdot)$ is the trained model and $y$ is the target class, typically different from the true class:
\begin{equation}
    \underset{\delta \in \Delta}{\text{minimize}} \, \ell(h_{\theta}(x + \delta), y)
    \label{equation: attack_obj}
\end{equation}

subject to $\Delta = \{ \delta : \| \delta \|_{\infty} \leq \epsilon \}$, where $\epsilon$ is the attack budget. However, using machine learning models to monitor contactless vital signs primarily falls under the realm of regression tasks where the labels are continuous values. When dealing with radar signals, it is essential to note that $x$ in Equation \ref{equation: attack_obj} represents radargrams rather than images. To generate adversaries for vital sign monitoring, we propose additional conditions that are imposed on the potential perturbation $\delta$. Here are two key design considerations for optimizing perturbations:

\begin{itemize}
\item \textbf{Localized Perturbations:} Rather than applying perturbations to the entire image or radargram, we advocate adding $\delta$ exclusively to the range bins where the target of interest is situated. For instance, if the target resides on the $n^{th}$ range bin, then the vital sign would predominantly span across the neighboring range bins of $n$. Consequently, perturbations should also be introduced to these pertinent range bins. This design consideration is ensured in the algorithm's process-step 4 through $d_j$ in Equation \ref{eqn:delta_i,j,k}, specifically through the \textit{offsets} parameter in the expression of $d_j$ as shown in Equation \ref{eqn: offset_equation}.

\item \textbf{Structured Perturbations:} The aforementioned approach of perturbation addition in Equation \ref{equation: attack_obj} operates on a per-pixel or per-element basis within the radargram, lacking structure and specificity tailored to deceive vital sign estimation models. As an alternative strategy, we propose the introduction of a periodic signal, specifically a sinusoid, atop the genuine periodic vital sign signature. This augmentation aims to ensure that the perturbation sinusoid is just enough to overwhelm the original vital sign, rendering it challenging for the vital sign estimation model to accurately predict true heart rate. This design consideration is realized in the algorithm introducing sinusoid with frequency $f_{osc}$ in the first term of $d_j$ as shown in Equation \ref{eqn: offset_equation}.

\end {itemize}

Our proposed Algorithm \ref{alg:sinusoidal_attack} incorporates the creation of a structured, localized perturbation guided by learned parameters through gradient optimization, reflecting the nature of the vital signature. The following equation represents the proposed defense objective, where $y$ is the target heart rate, typically away from the true heart rate:
\begin{equation}
    \underset{\delta \in \Delta}{\text{minimize}} \, \ell(h_{\theta}(x_{i,j} + \delta_{i,j,k} ), y)
\end{equation}
In our case, $\delta_{i,j,k} $ is a function of optimized frequency and spatial amplitude of the perturbation sinusoid in the radargram, i.e. \(\delta_{i,j,k}: f \mapsto f(A_{\text{opt}}, f_{\text{opt}})\)



Our attack budget is not defined by a single parameter but rather by a set of constraints. Specifically, it is a combination of several factors where the constraints are represented as $\Delta = \{ \delta: \delta_A \leq \epsilon_A, \delta_f \leq \epsilon_f \}$.


\begin{algorithm}[ht]
\caption{Sinusoidal Defense Algorithm}
\label{alg:sinusoidal_attack}
 
 \textbf{Input:} 
    \begin{itemize}
        \item $h_\theta(\cdot )$: Target (trained) model
        \item $x \in \mathbb{R}^{M \times N}$: Original radargram
        \item $y$: Targeted HR (away from true HR)
        \item $\alpha$: Step size
        \item $T$: Number of iterations
        \item $L(\cdot)$: Loss function 
    \end{itemize}
    
\textbf{Output:} 
    \begin{itemize}
        \item Optimized frequency $f_{\text{opt}}$ and spatial amplitude $A_{\text{opt}}$
        \item Perturbed radargram $x' \in \mathbb{R}^{M \times N}$
    \end{itemize}

\textbf{Process:}
\begin{algorithmic}[1]
\STATE Initialize $x' \gets x$
\STATE Initialize frequency estimate $f_{\text{opt}} \gets \text{random number} \in [f_{\min}, f_{\max}]$
\FOR{$t = 1$ \TO $T$}
    \STATE Generate synthetic radargram perturbation with $d_j$ defined by Equation \ref{eqn: offset_equation}:
    \small 
    \begin{align}
\delta_{i,j,k}(A_{opt}, f_{\text{opt}}) &= \exp\left(-\left(\frac{t_k - d_j}{\omega_0 / T_s}\right)^2\right) \cdot \nonumber \\
& \quad \exp\left(i \cdot 2\pi \cdot f_{\text{0}} \cdot T_s (t_k - d_j)\right)
\label{eqn:delta_i,j,k}
\end{align}
\vspace{-0.25cm}
    \STATE Add the perturbation to the original radargram:
    \[
    x' \gets x + \delta_{i,j}(A_\text{opt}, f_{\text{opt}})
    \]
    \vspace{-0.25cm}
    \STATE Compute the gradient of the loss w.r.t. $f_{\text{opt}}$ and $A_{\text{opt}}$:
    \[
    G_f \gets \nabla_{f_{\text{opt}}} L(h_\theta(x'), y)
    \hspace{5pt} \& \hspace{5pt}
    G_A \gets \nabla_{A_{\text{opt}}} L(h_\theta(x'), y)
    \]
    \vspace{-0.25cm}
    \STATE Update the estimated frequency using gradient descent:
    \[
    f_{\text{opt}} \gets f_{\text{opt}} - \alpha \cdot G_f
    \hspace{5pt} \& \hspace{5pt}
    A_{\text{opt}} \gets A_{\text{opt}} - \alpha \cdot G_A
    \]
    \vspace{-0.25cm}
    \STATE Clip the updated frequency and spatial amplitude to ensure it stays within predefined bounds:
    \[
    f_{\text{opt}}\gets \text{clip}(f_{\text{opt}}, f_{\min}, f_{\max}) \hspace{3pt} \& \] 
    \[
    A_{\text{opt}}\gets \text{clip}(A_{\text{opt}}, A_{\min}, A_{\max})
    \]
    \vspace{-0.35cm}
\ENDFOR
\STATE \textbf{return}  Optimized $f_{\text{opt}}$, $A_{\text{opt}}$, and  $x'$
\end{algorithmic}
\end{algorithm}
\subsection{Constraint on Frequency:}
For heart rate estimation, we should ensure that it is recommended that the deviation $\delta_f$ from the true frequency be kept within $\epsilon_{hr}$, ensuring that the estimated heart rate falls within the physiological range of 50 to 100 beats per minute, indicative of normal human heart rate. Any data falling outside this range should be classified as noise and disregarded by the system.

    


\subsection{Constraint on Spatial Amplitude/ Span:}
The spatial amplitude of the perturbation signal, also known as the perturbation sinusoid, should be kept within the target's occupancy range bins, such that $\delta_{A} \leq \epsilon_{A}$, where $\mid \epsilon_A \mid$ is $25$ range bins. This limitation arises from human targets typically occupying an average of $46$ cm, which is around $50$ range bins at specific time intervals, given the scale of $9$ mm per range bin in P440 UWB radar \cite{P440}.
\\

The algorithm optimizes both frequency $f_{\text{opt}}$ and spatial amplitude $A_{\text{opt}}$ of the sinusoidal perturbation based on the above constraints to ensure the attack remains within the physiological and spatial limits defined for heart rate estimation and radargram occupancy.

\section{Hardware and Measurement Setup}
\subsection{UWB Radar Setup}
We utilize a monostatic time domain Ultra-WideBand (UWB) Impulse Radar P440 \cite{P440} with time windowing capabilities for sensing vital signs. It operates from 3.1 to 4.8 GHz frequency centering at 4.3 GHz. 

\subsection{Programmable Servo Motor}
\begin{figure}[htb!]
    \centering
    \includegraphics[width=0.47\textwidth]{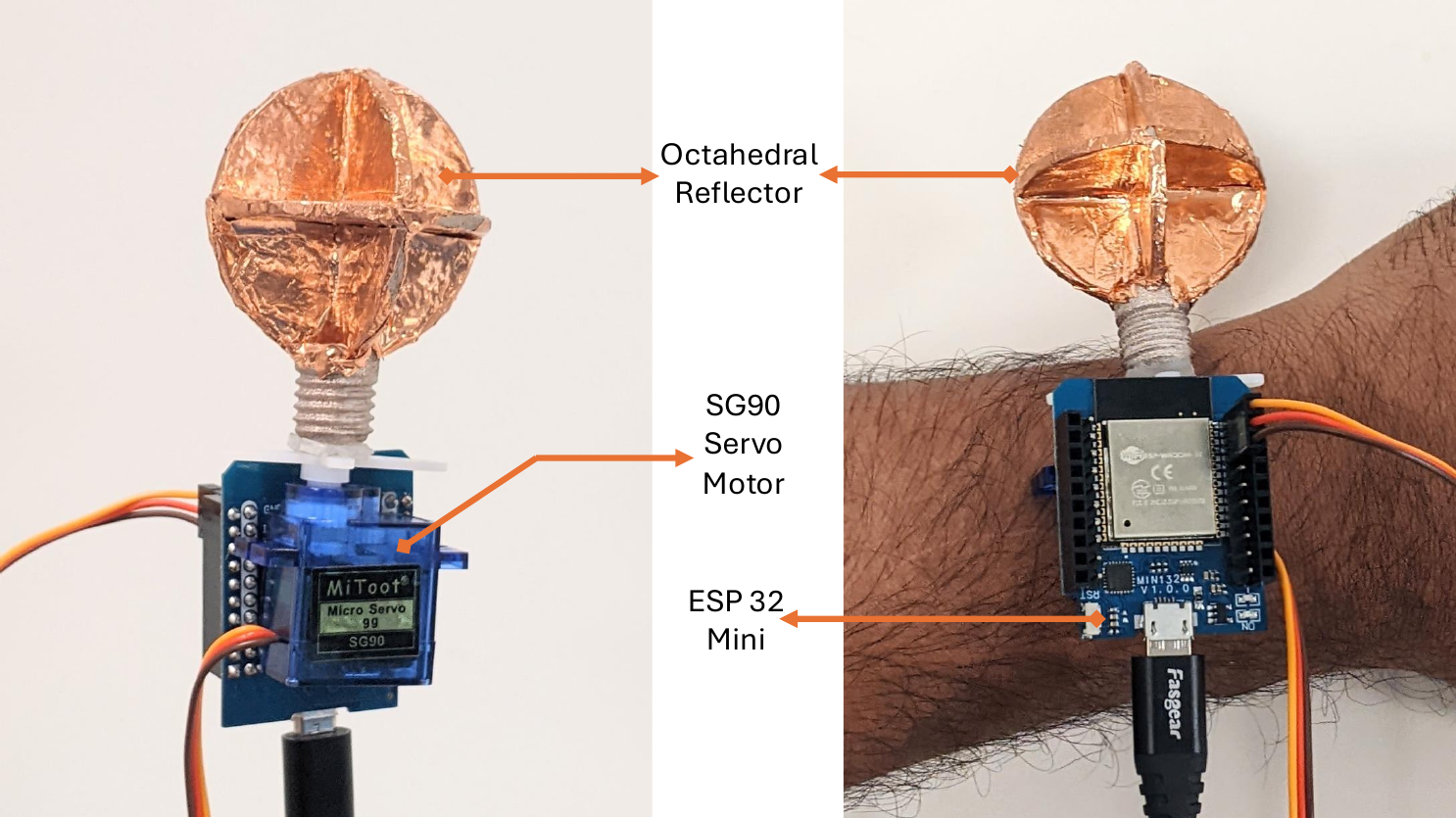} 
    \caption{A programmable servo motor paired with a ESP 32 Mini. A 3D-printed octahedral reflector wrapped with copper tape is attached as a load to the motor to enhance reflectivity and increase the signal-to-noise ratio (SNR).}
    \label{fig: servo_motor}
\end{figure}

As an anti-sensing mechanism, we propose using an off-the-shelf SG90 servo motor programmed via an ESP32 WROOM Mini microcontroller to generate variable frequency motion tailored to specific requirements. By optimizing the perturbation $f_\text{opt}$ within predefined constraints (normal heart rates range from $50$ bpm to $100$ bpm), this setup allows for precise, reproducible frequency motion that can effectively counteract radar sensing mechanisms. A 3D-printed octahedral reflector, with a diameter of $4.3$ cm and wrapped in copper tape, is attached to the motor to improve reflectivity and enhance the signal-to-noise ratio (SNR). Figure \ref{fig: servo_motor} illustrates the setup of the anti-sensing device, which is compact enough to be worn on the wrist.

\begin{table*}
\centering
\begin{tabular}{|c| c| c| c|c|}
\hline
\textbf{Dataset} & \textbf{Setting} & \textbf{HR Model} & \textbf{MAE (Without Anti-Sensing)} & \textbf{MAE (With Anti-Sensing)} \\
\hline
\multirow{3}{*}{Tasnim et al. \cite{oshim2022eulerian}} & \multirow{3}{*} {Sleep Lab} & ResNet - 18 \cite{saeed2021portable} & 2.67 bpm & 5.35 bpm  \\
 &  & ResNet - 50 \cite{saeed2021portable}& 2.37 bpm & 5.63 bpm  \\
 &  & CNN 1D+2D \cite{CNN1D2D} & 5.63 bpm & 12.91 bpm  \\
 &  & Vision Transformer (ViT) \cite{dosovitskiy2020image} & 3.28 bpm & 9.35 bpm  \\
\hline
\end{tabular}
\caption{HR Sensing Model Performance with and without Anti-Sensing on Sleep Dataset}
\label{Tab: MAE}
\end{table*}




\section{Results}

\subsection{Anti-Sensing on Deep Learning-based Vital Sign Estimation Models}



\begin{figure}[htb!]
    \centering
    \begin{subfigure}[b]{0.22\textwidth}
        \centering
        \includegraphics[width=\textwidth]{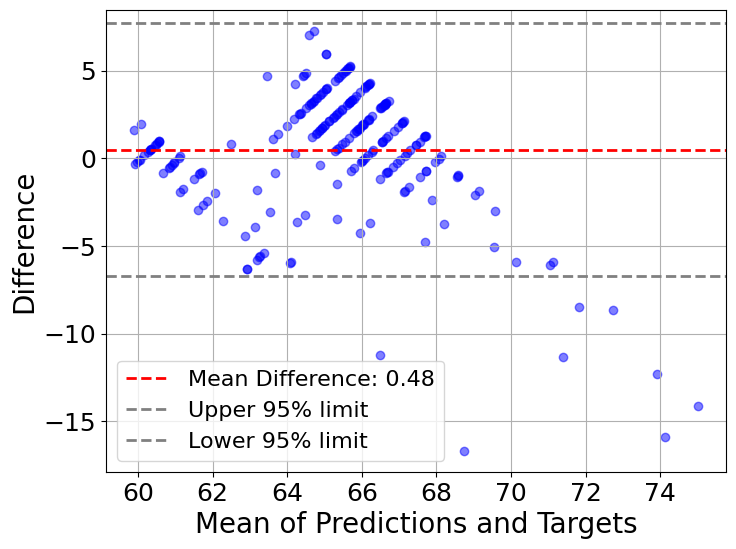}
        \caption{W/o Anti-Sensing (Resnet-18)}
        \label{fig: svs_before_attack_18}
    \end{subfigure}
    \hfill
    \begin{subfigure}[b]{0.22\textwidth}
        \centering
        \includegraphics[width=\textwidth]{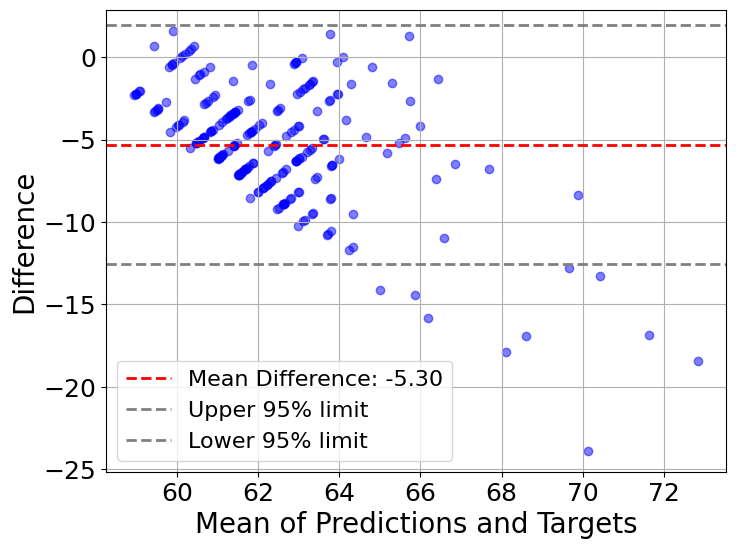}
        \caption{With Anti-Sensing (Resnet-18)}
        \label{fig: svs_after_attack_18}
    \end{subfigure}
    
    \vspace{0.5cm} 
    
    \begin{subfigure}[b]{0.22\textwidth}
        \centering
        \includegraphics[width=\textwidth]{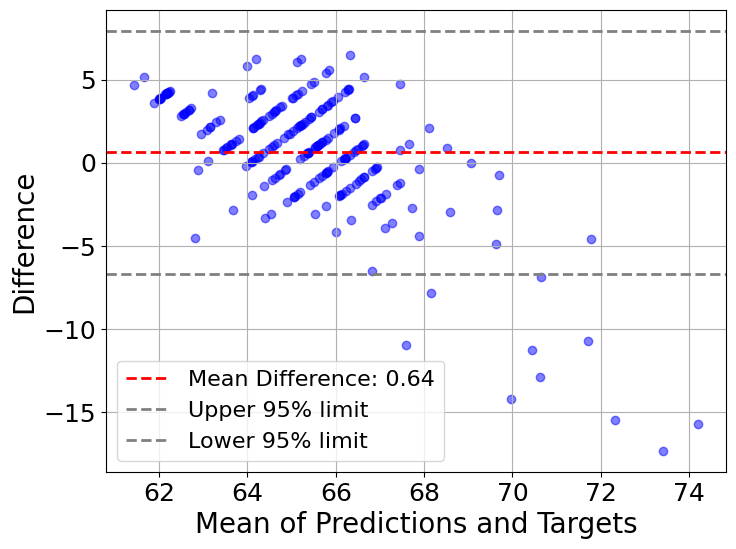}
        \caption{W/o Anti-Sensing (Resnet-50)}
        \label{fig: svs_before_attack_50}
    \end{subfigure}
    \hfill
    \begin{subfigure}[b]{0.22\textwidth}
        \centering
        \includegraphics[width=\textwidth]{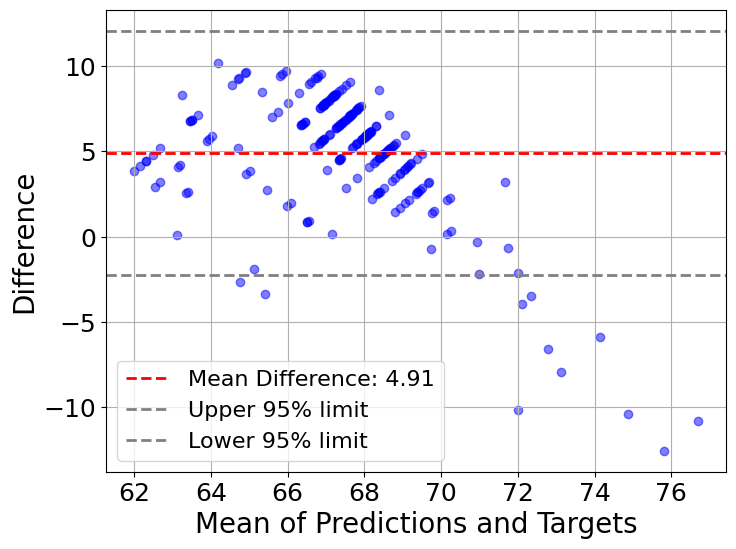}
        \caption{With Anti-Sensing (Resnet-50)}
        \label{fig: svs_after_attack_50}
    \end{subfigure}

    \vspace{0.5cm} 
    
    \begin{subfigure}[b]{0.22\textwidth}
        \centering
        \includegraphics[width=\textwidth]{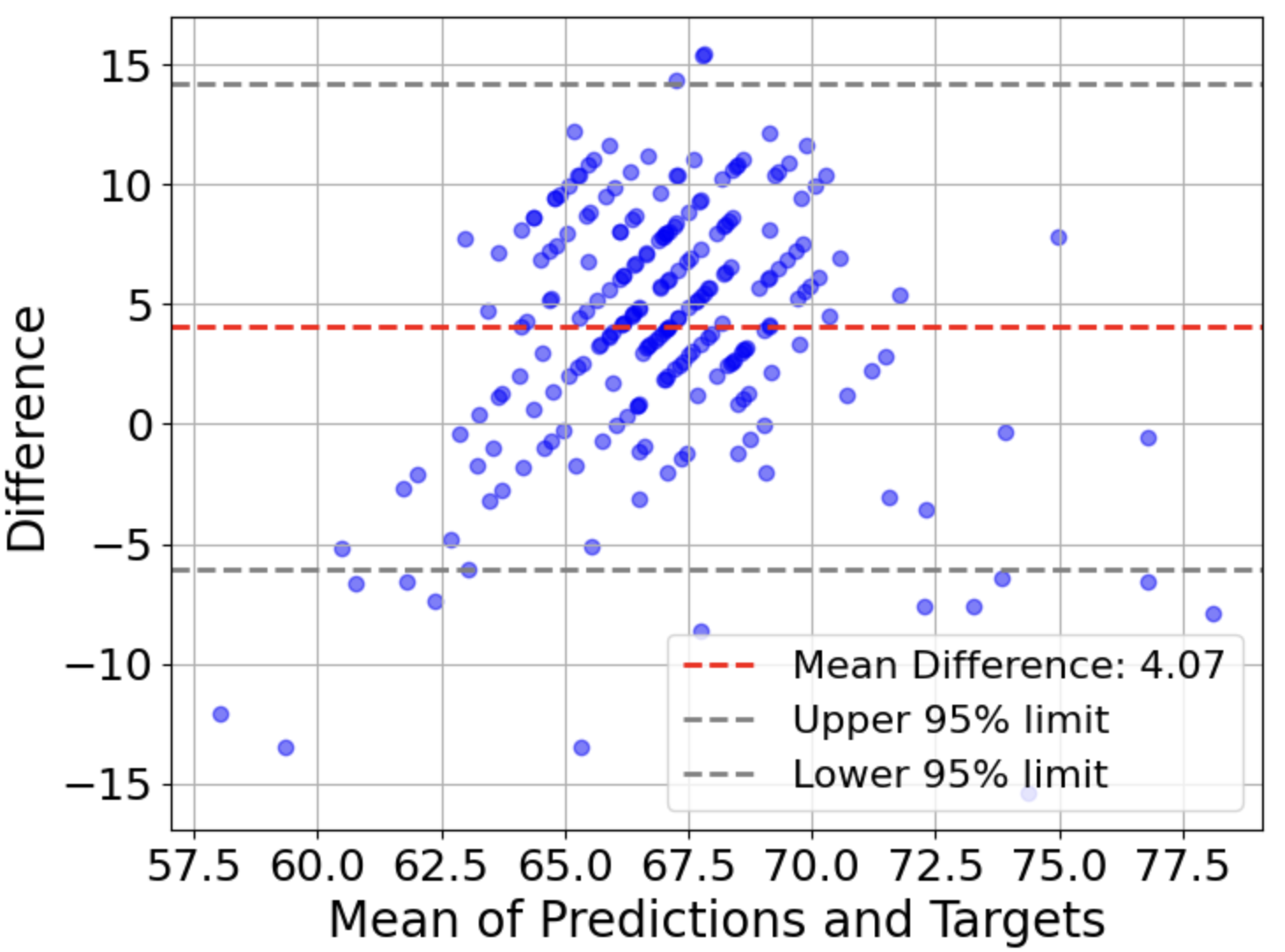}
        \caption{W/o Anti-Sensing (CNN 1D+2D)}
        \label{fig: svs_before_attack_50}
    \end{subfigure}
    \hfill
    \begin{subfigure}[b]{0.22\textwidth}
        \centering
        \includegraphics[width=\textwidth]{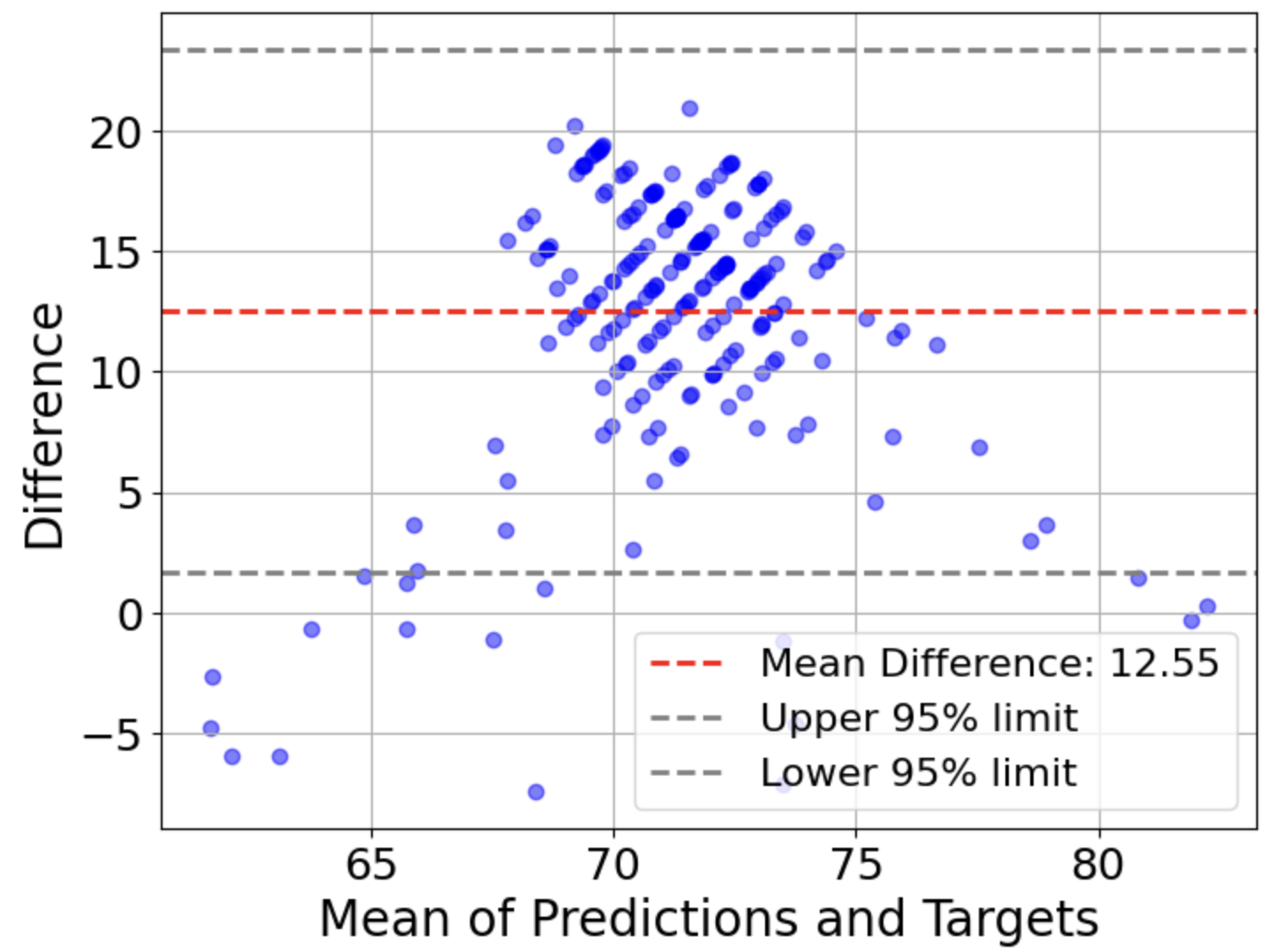}
        \caption{With Anti-Sensing (CNN 1D+2D)}
        \label{fig: svs_after_attack_50}
    \end{subfigure}

    \vspace{0.5cm} 
    
    \begin{subfigure}[b]{0.22\textwidth}
        \centering
        \includegraphics[width=\textwidth]{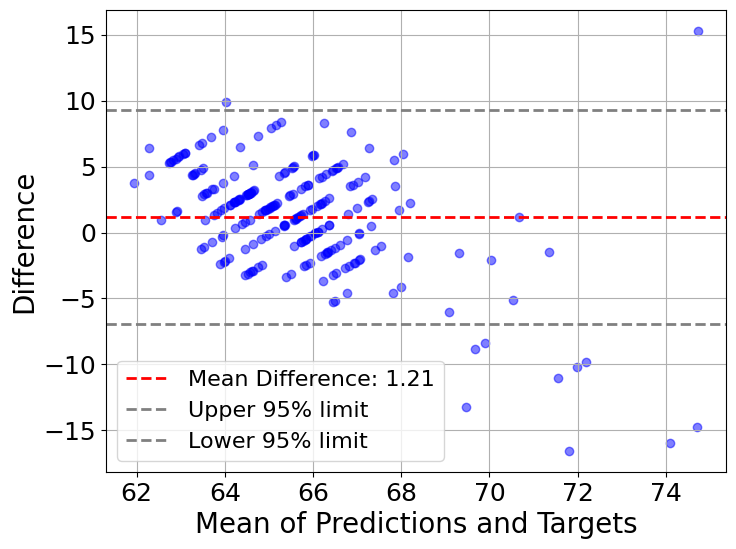}
        \caption{W/o Anti-Sensing (ViT)}
        \label{fig: svs_before_attack_50}
    \end{subfigure}
    \hfill
    \begin{subfigure}[b]{0.22\textwidth}
        \centering
        \includegraphics[width=\textwidth]{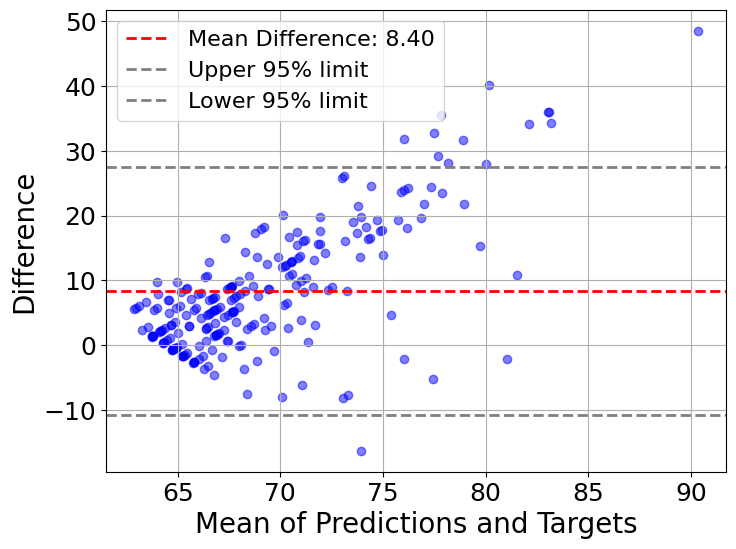}
        \caption{With Anti-Sensing (ViT)}
        \label{fig: svs_after_attack_50}
    \end{subfigure}
    
    \caption{Comparison of Bland Altman Plots from Resnet-18, Resnet-50, CNN 1D+2D, and  Vision Transformer (ViT) models for HR estimation without (left column) and with (right column) anti-sensing perturbation applied on sleep dataset.}
    \label{fig: bland_altman_comparison}
\end{figure}

\subsubsection{Dataset}

To validate our proposed anti-sensing defense algorithm, we used a sleep dataset \cite{oshim2022eulerian} collected in a sleep laboratory. Two participants underwent overnight full polysomnography (Siesta, COMPUMEDICS), which included electrocardiography and respiratory inductance plethysmography to measure chest and abdominal wall motion along with simultaneous contactless UWB radar data collection.

\subsubsection{Models}

We employed pre-trained ResNet-18, ResNet-50 \cite{saeed2021portable}, and Vision Transformer (ViT) \cite{dosovitskiy2020image} models and fine-tuned them on the sleep dataset for regression task of heart rate estimation. Additionally, we used a CNN-based model \cite{CNN1D2D} that combines both 1D and 2D signal extraction approaches.

Figure \ref{fig: bland_altman_comparison} shows the Bland-Altman plots comparing the pre-trained and fine-tuned ResNet-18, ResNet-50, CNN 1D+2D, and Vision Transformer (ViT) models on sleep clinic data with and without anti-sensing perturbations. Without anti-sensing, the mean differences were $0.48, 0.64, 4.07$, and $1.21$ for ResNet-18, ResNet-50, CNN 1D+2D, and ViT respectively, indicating a close match between predictions and ground truth. However, with anti-sensing applied, the mean differences increased to $-5.30, 4.91, 12.55$, and $8.40$, respectively, highlighting the algorithm's effectiveness in disrupting model accuracy.

\begin{figure}[htb!]
    \centering
    \begin{subfigure}[b]{0.22\textwidth}
        \centering
        \includegraphics[width=\textwidth]{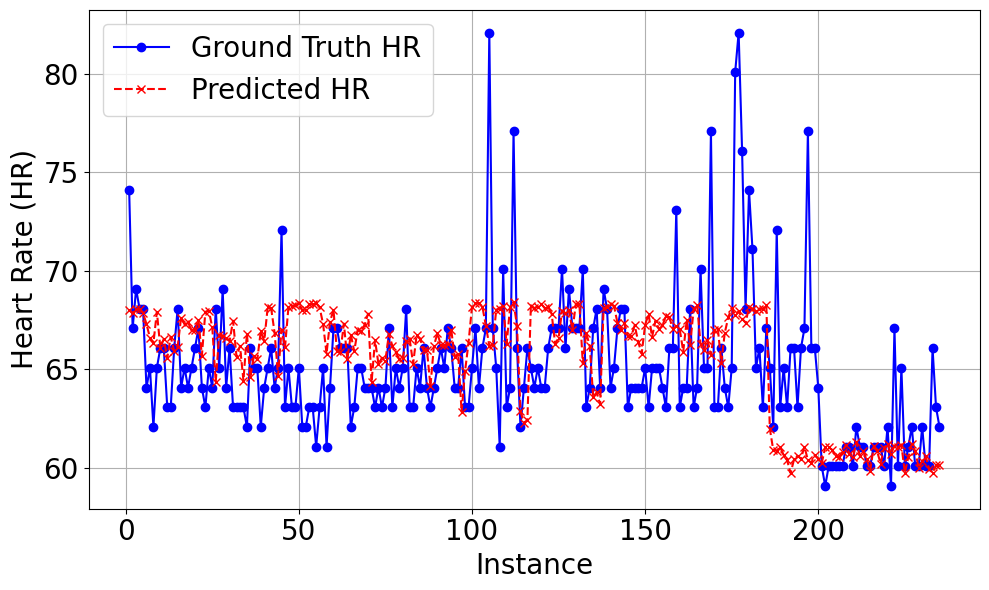}
        \caption{W/o Anti-Sensing (Resnet-18)}
        \label{fig: resnet18_before_attack}
    \end{subfigure}
    \hfill
    \begin{subfigure}[b]{0.22\textwidth}
        \centering
        \includegraphics[width=\textwidth]{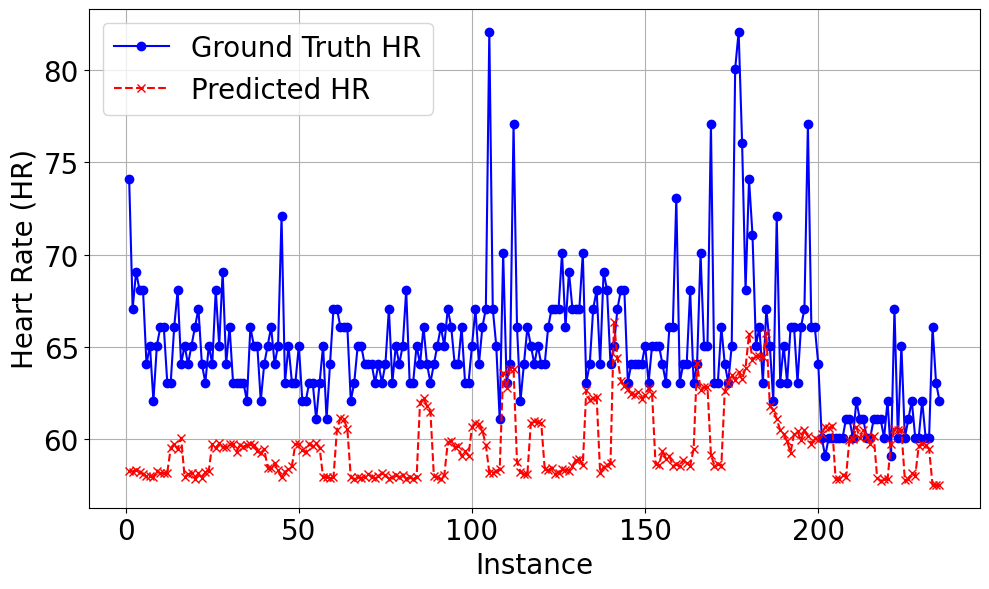}
        \caption{With Anti-Sensing (Resnet-18)}
        \label{fig: resnet18_after_attack}
    \end{subfigure}
    
    \vspace{0.25cm} 
    
    \begin{subfigure}[b]{0.22\textwidth}
        \centering
        \includegraphics[width=\textwidth]{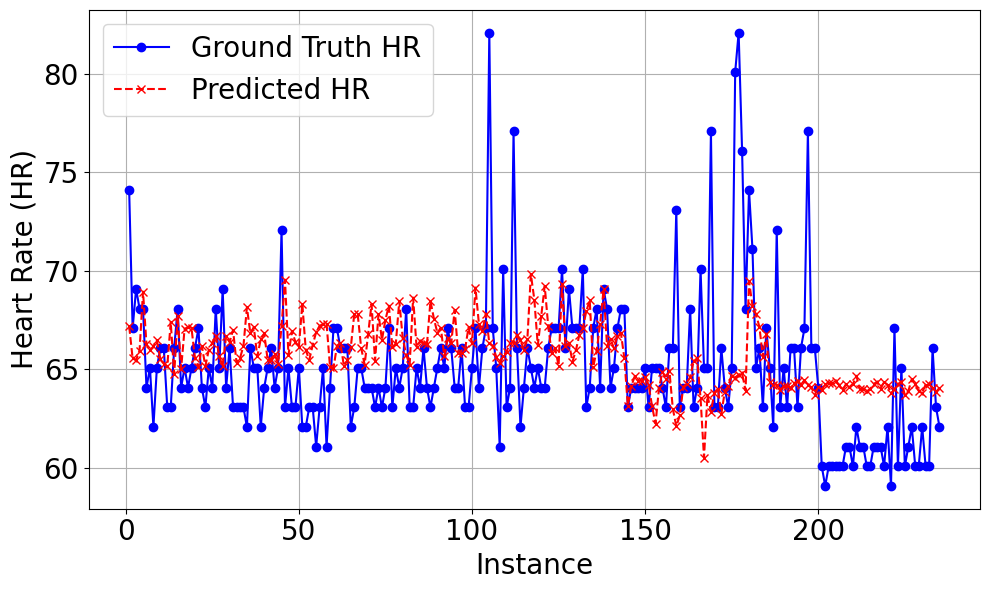}
        \caption{W/o Anti-Sensing (Resnet-50)}
        \label{fig: resnet50_before_attack}
    \end{subfigure}
    \hfill
    \begin{subfigure}[b]{0.22\textwidth}
        \centering
        \includegraphics[width=\textwidth]{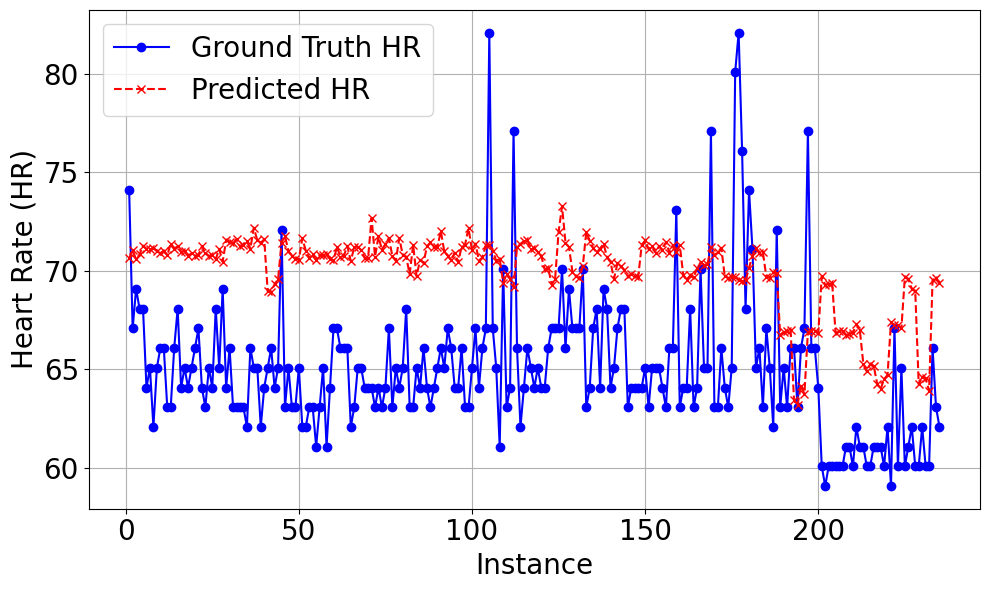}
        \caption{With Anti-Sensing (Resnet-50)}
        \label{fig: resnet50_after_attack}
    \end{subfigure}
    
    \vspace{0.25cm} 
    
    \begin{subfigure}[b]{0.22\textwidth}
        \centering
        \includegraphics[width=\textwidth]{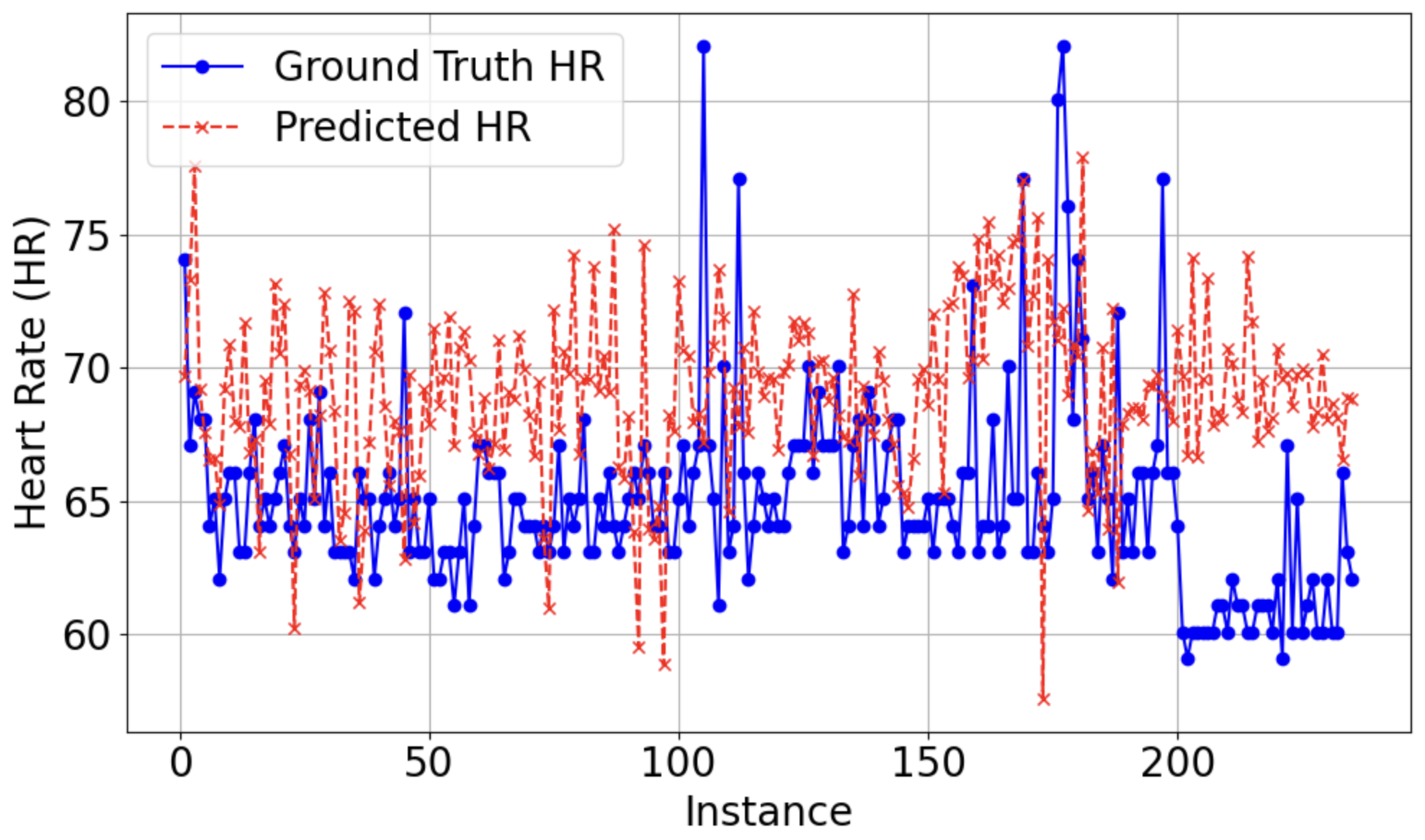}
        \caption{W/o Anti-Sensing (CNN 1D+2D)}
        \label{fig: resnet50_before_attack}
    \end{subfigure}
    \hfill
    \begin{subfigure}[b]{0.22\textwidth}
        \centering
        \includegraphics[width=\textwidth]{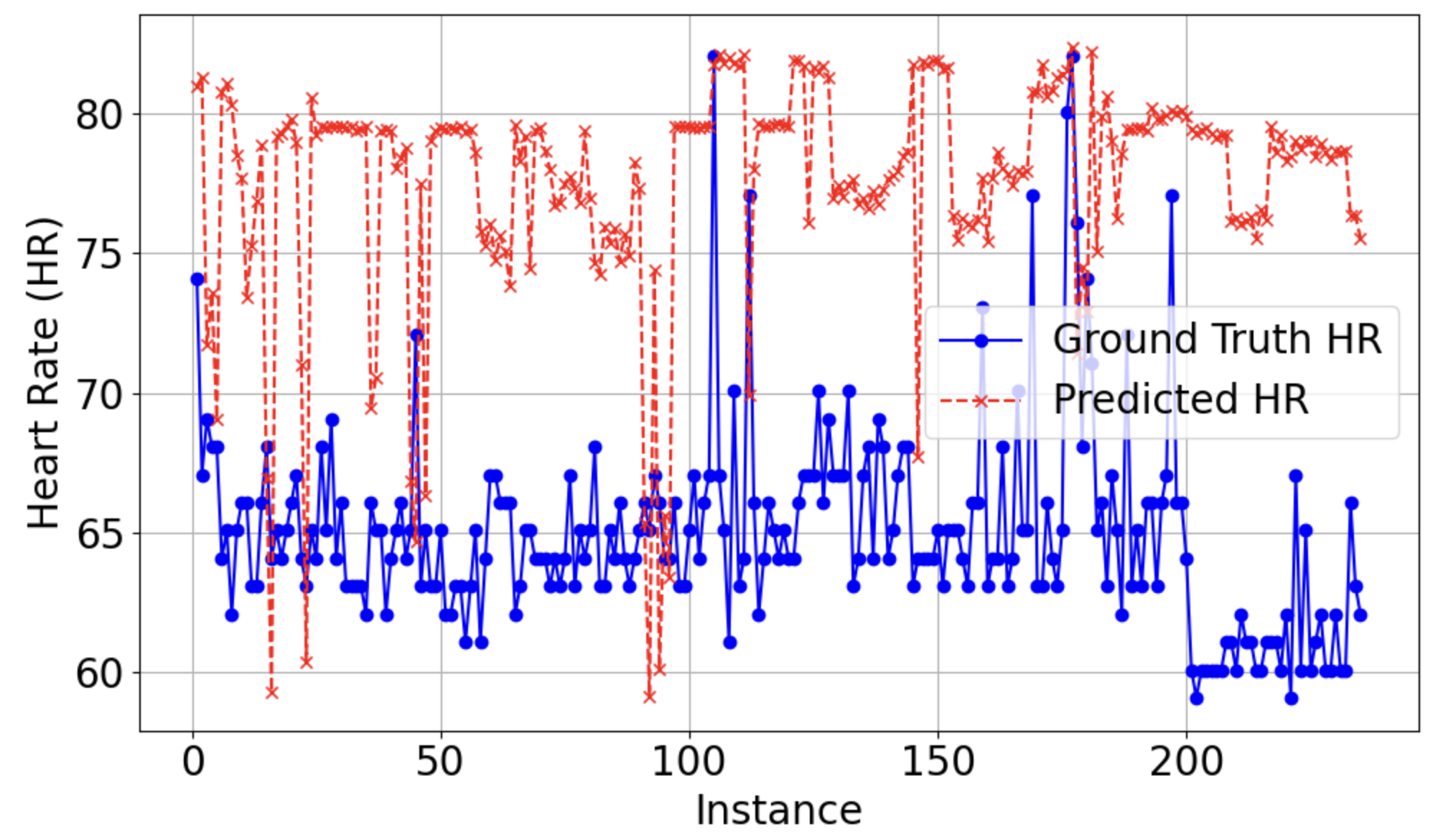}
        \caption{With Anti-Sensing (CNN 1D+2D)}
        \label{fig: resnet50_after_attack}
    \end{subfigure}

    \vspace{0.25cm} 
    
    \begin{subfigure}[b]{0.22\textwidth}
        \centering
        \includegraphics[width=\textwidth]{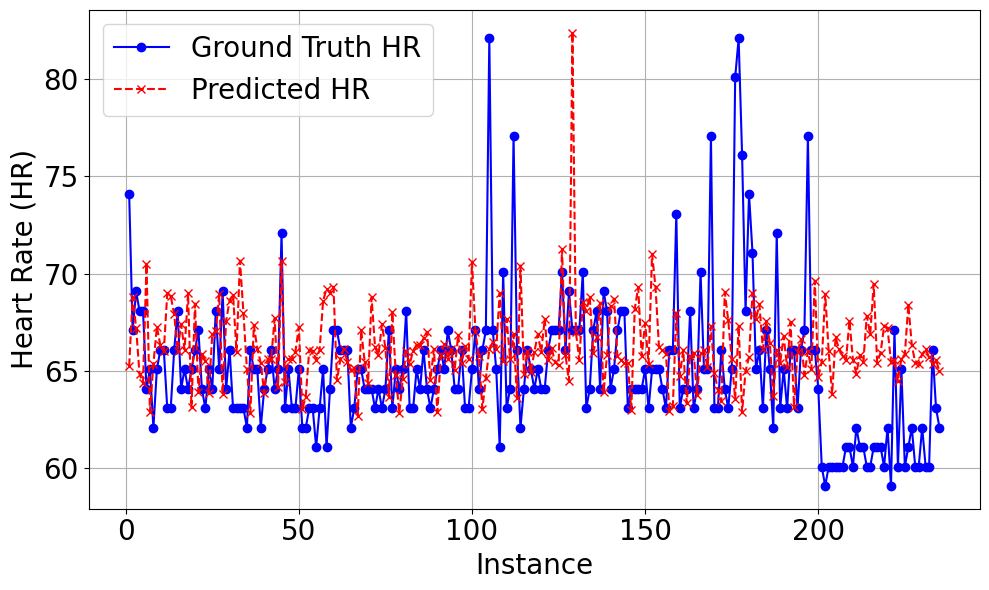}
        \caption{W/o Anti-Sensing (ViT)}
        \label{fig: resnet50_before_attack}
    \end{subfigure}
    \hfill
    \begin{subfigure}[b]{0.22\textwidth}
        \centering
        \includegraphics[width=\textwidth]{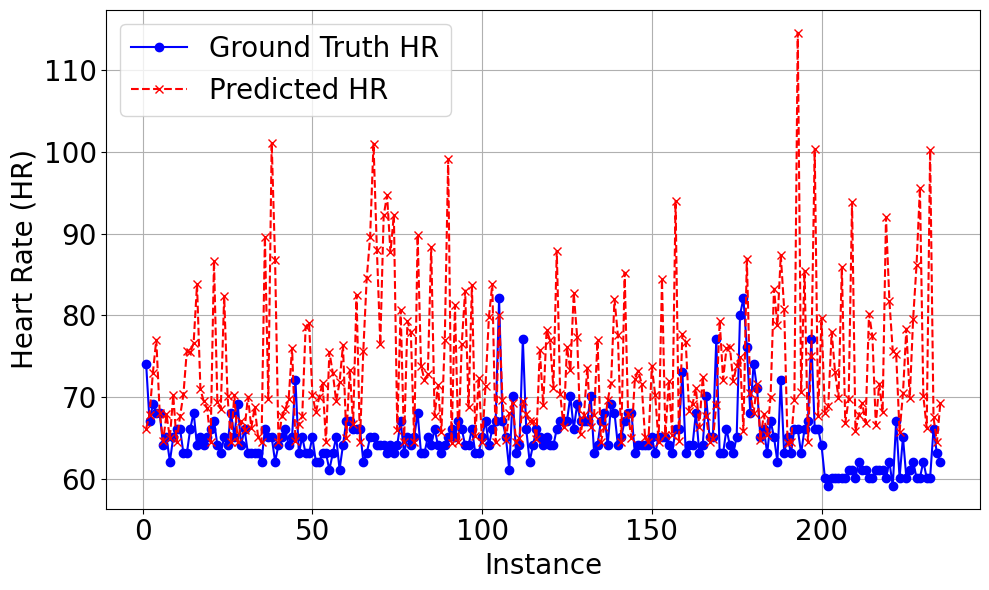}
        \caption{With Anti-Sensing (ViT)}
        \label{fig: resnet50_after_attack}
    \end{subfigure}
    \caption{Comparison of Resnet-18, Resnet-50, CNN 1D+2D, and  Vision Transformer (ViT) model performance on HR prediction without (left column) and with (right column) anti-sensing perturbation applied on sleep dataset.}
    \label{fig: resnet_comparison}
\end{figure}

Figure \ref{fig: resnet_comparison} further illustrates the growing disparity between predicted and actual heart rates under anti-sensing, while Table \ref{Tab: MAE} shows a significant increase in MAE after the attack: $2.68$ bpm for ResNet-18, $3.26$ bpm for ResNet-50, $7.28$ bpm for CNN 1D+2D, and $6.07$ bpm for ViT.

\subsection{Evaluating our Anti-Sensing Device}

\begin{figure}
  \includegraphics[width=0.5\textwidth]{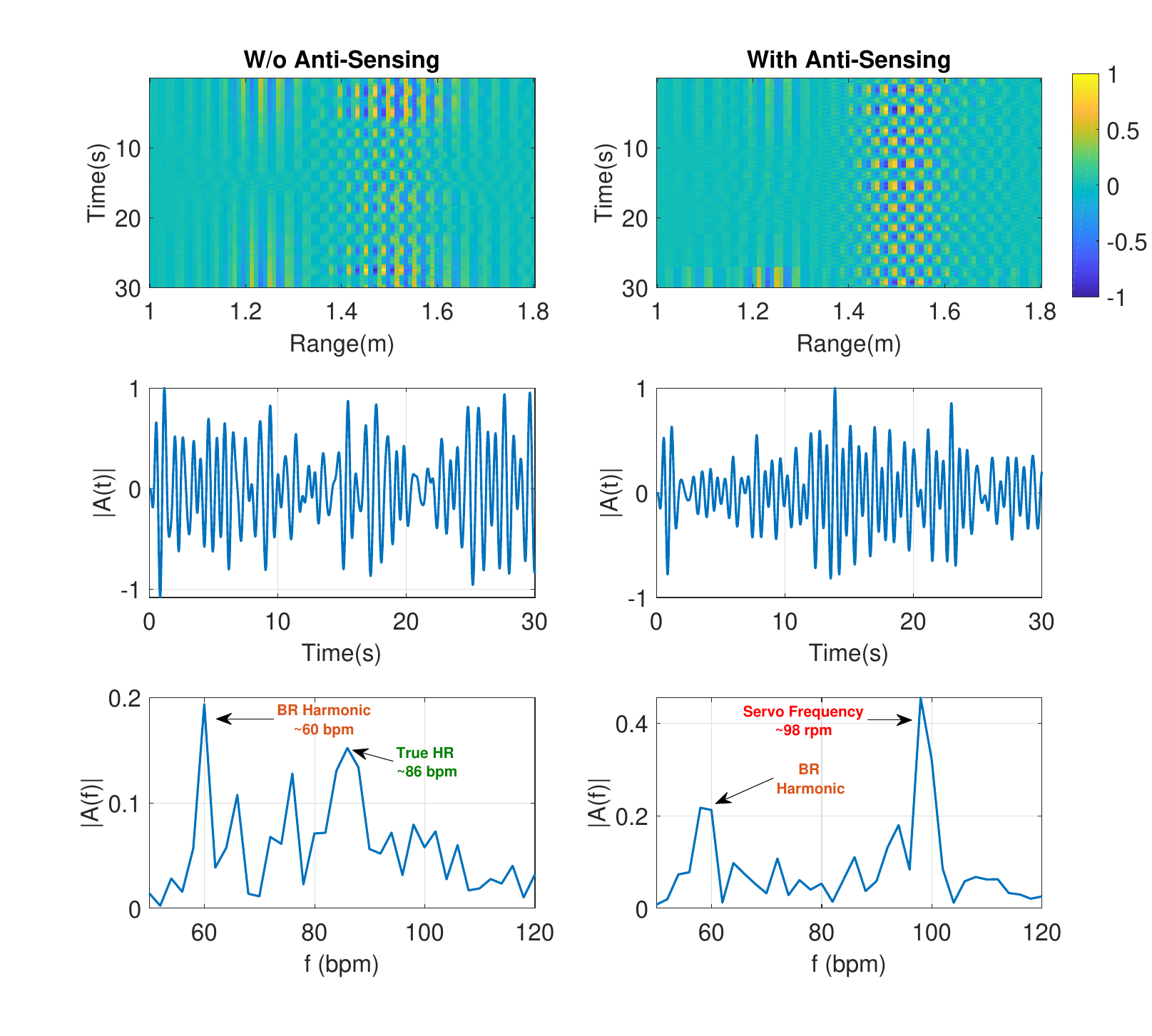}
  \caption{Sample comparison of HR measurement with (right column) and without (left column) anti-sensing motor running at 98 RPM in front of the participant. The top row displays the corresponding 2D radargrams, the middle row shows the 1D extracted signals, and the bottom row features the FFT plots.}
  \label{fig: real_servo_comparison}
\end{figure}

Figure \ref{fig: real_servo_comparison} shows a comparative analysis of a sample human heart rate measurement with and without the anti-sensing motor running at $98$ RPM positioned in front of the participant. Without anti-sensing (bottom left), the FFT plot displays the ground truth heart rate of 86 bpm as a prominent peak. Conversely, the FFT plot for the case with anti-sensing (bottom right) shows the anti-sensing motor's frequency as the highest peak, demonstrating the impact of the anti-sensing motor on the measurement. The harmonics of the breathing rate (BR) at $60$ bpm are evident in each FFT, as the participant's breathing rate was $20$ bpm.

\begin{table} [htb!]
\centering
\begin{tabular}{|c| c|c|}
\hline
 \textbf{HR Model} & \textbf{MAE (W/o Motor)} & \textbf{MAE (With Motor)} \\
\hline
 FFT & 2.42 bpm & 8.17 bpm  \\
 ResNet - 50 & 2.93 bpm & 7.23 bpm  \\
 CNN 1D+2D & 9.24 bpm & 17.56 bpm  \\
\hline
\end{tabular}
\caption{Performance on Real Anti-Sensing motor attached to the wrist.}
\label{Tab: MAE_Real}
\end{table}

To evaluate the performance of our anti-sensing device, we collected data from five individuals wearing it on their wrists, with the wrist positioned close to the chest. The ground truth was collected with a Galaxy smart watch. For each participant, we first executed the sinusoidal defense as depicted in Algorithm \ref{alg:sinusoidal_attack} for a specific heart rate (HR) model to determine the optimal servo frequency and spatial amplitude. Subsequently, the ESP32 microcontroller was programmed to rotate the servo at the optimized frequency. The spatial amplitude of the system was adjusted by modifying the arm length to which the octahedral reflector was mounted.

Table \ref{Tab: MAE_Real} shows the impact of the anti-sensing motor on the mean absolute error (MAE) of three heart rate (HR) models. For the FFT model, the MAE increased from $2.42$ bpm without the motor to $8.17$ bpm with the motor. Similarly, the ResNet-50 model experienced an increase in MAE from $2.94$ bpm without the motor to $7.23$ bpm with the motor. The CNN 1D+2D model showed a notable rise in MAE from $9.24$ bpm without the motor to $17.56$ bpm with the motor. The rest of the models perform very poorly, even without anti-sensing; hence, the results are not included. Overall, introducing the anti-sensing motor led to a degradation in performance for all models, with the CNN 1D+2D model exhibiting the most substantial increase in error. The varying degrees of performance degradation across different heart rate models can be attributed to differences in model architectures and the optimized perturbation frequency, which varies across participants.



\section{Conclusion and Future Work}
This paper presents Anti-Sensing, a novel defense against unauthorized radar-based heart rate sensing. By introducing physically realizable perturbations via a wearable device, we disrupted radar sensing models, leading to inaccurate heart rate estimations and enhanced privacy protection. Our gradient-based algorithm optimized device oscillations within physiological limits, with experiments validating its effectiveness. Given the growing reliance on radar for human sensing in robotics, this research is crucial for ensuring privacy and security in next-generation robotic systems.

While this work establishes the foundation for physical anti-sensing in the radar domain, future efforts will extend the technique to more complex tasks, such as gesture recognition, and explore multi-modal defenses. In future, we aim to refine the optimization of the perturbation mechanisms to be real-time, lightweight, compact, and potentially battery-free, further reducing user burden.

\bibliographystyle{IEEEtran}
\bibliography{arxiv_main}

\end{document}